\documentclass[conference]{IEEEtran}
\IEEEoverridecommandlockouts
\usepackage{cite}
\usepackage{amsmath,amssymb,amsfonts}
\usepackage{algorithmic}
\usepackage{graphicx}
\usepackage{textcomp}
\usepackage{xcolor}
\usepackage{hyperref}

\usepackage[most]{tcolorbox}
\usepackage{enumitem}
\usepackage{eurosym}

\def\BibTeX{{\rm B\kern-.05em{\sc i\kern-.025em b}\kern-.08em
    T\kern-.1667em\lower.7ex\hbox{E}\kern-.125emX}}
\begin{document}

\title{Advancing a Responsible Future Quantum Internet
\thanks{We acknowledge funding from Quantum Delta NL through The Centre for Quantum \& Society and the Dutch Research Council (NWO) through the project “QuTech Part II Applied-oriented research” (project number 601.QT.001) }}

\author{\IEEEauthorblockN{1\textsuperscript{st} K.L. van der Enden}
\IEEEauthorblockA{\textit{QuTech and Kavli Institute of Nanoscience} \\
\textit{Delft University of Technology}\\
Delft, The Netherlands \\
\href{mailto:publications@interseqt.nl}{publications@interseqt.nl}}
\and
\IEEEauthorblockN{2\textsuperscript{nd} G. Profitiliotis}
\IEEEauthorblockA{\textit{Centre for Quantum \& Society} \\
\textit{Quantum Delta NL}\\
Delft, The Netherlands \\
}
\and
\IEEEauthorblockN{3\textsuperscript{rd} D. Croese}
\IEEEauthorblockA{\textit{Centre for Quantum \& Society} \\
\textit{Quantum Delta NL}\\
Delft, The Netherlands}
}

\maketitle
\begin{abstract}
We evaluate the status of the development of a responsible future quantum internet (QI). Through horizon scanning, domain expert trend analysis and guided workshops, we present a desired future (DF) conceptualized by stakeholders within the scope of the ethical, legal, societal aspects (ELSA) and policy implications (ELSPI) of the QI. We examine the alignment of the present situation and the DF of the QI ELSPI to the ideals of the `ten principles for responsible quantum innovation' developed by [Mauritz Kop \textit{et al.} 2024 \textit{Quantum Sci. Technol.} 9 035013]. Most principles in the DF are well aligned to the ideal, except for the misalignment in `intellectual property' (IP) and `dual use', revealing the precarious balance of well intended policy suggestions and effective outcomes. Additionally, there is an overemphasis placed on the principal of societal relevance in the DF, risking overseeing other principles in the future steering of the ELSPI. The present situation is in moderate alignment with the principles, however trending to misalignment on IP and international collaboration due to QI commercialization and the push for geopolitical sovereignty. For continued success of a responsible quantum internet, we recommend further investigation on the prevention of dual use quantum internet applications, closer involvement of commercial entities in ELSPI ideation, continued recognition of base-layer technology research and stakeholder education on QI applications.
\end{abstract}

\section{Introduction}
The development of quantum technologies has seen great advancements in the last decade driven by the potential to revolutionize computation, communication and sensing~\cite{Kimble2008, Degen2017, Bova2021, Bayerstadler2021}. 
 
In parallel to quantum computing, the development of technology that allows quantum computers to share quantum information is now more relevant than ever. This `quantum internet' (QI) is making its way to become a powerhouse in accelerating the pace to useful quantum computing by enabling distributed quantum computing, inherently secure communication and anonymous transmission~\cite{Wehner2018}. 

Although profoundly unique and useful from a security and computational perspective, the development of the QI equally poses risks for bad actors to operate with impunity, can cause international informational network isolation and inequality in (quantum) resource access. With the several levels of functionality of the QI rapidly increasing in Technology Readiness Level (TRL)~\cite{Wehner2018}, it is of importance to evaluate the ethical, legal, societal and policy implications (ELSPI) of a future QI~\cite{kop2023quantum}. Previous work has introduced guardrails to steer innovation without hindering its progress with the `ten principles for responsible quantum innovation'~\cite{Kop2024}.

In order to facilitate an ELSPI analysis of a future QI, it is necessary to actively consider such a future. Without consideration of the future, humans are bounded by the simulation heuristic, i.e. a difficult to imagine future tends to be seen as unlikely. Simultaneously, different stakeholders do not necessarily share the same mental model of the future, which is a breeding ground for unrecognized assumptions and biases. A way to counteract both challenges is to follow a process of collaborative foresight conceptualization~\cite{Fuller2009}.

To facilitate foresight ideation and co-create a so-called `desired future' (DF) for the QI, the Dutch ‘Centre for Quantum and Society’ sponsored the transdisciplinary strategic foresight project 'Scenarios for Quantum Networks'\footnote{Project conducted by Dr. G. Profitiliotis in 2023-2024.}. For this project, the Centre organized a series of workshops where a diverse group of representatives of academic, industrial, governmental and civil society stakeholders were invited to participate in a collaborative scenario development process of the QI in 2050\footnote{The timeline of 2050 is set to symbolically represent a long-term future.}. In total more than 50 people participated, including a 10 person core team of experts from a diverse background within the quantum ecosystem to be present along the entire process of the DF creation.  

In this paper we evaluate the status of the development of a responsible future QI. We utilize the `ten principles for responsible quantum innovation' and evaluate the status quo and DF of the QI along the principles' ideals. First we will discuss the process towards the creation of this DF in Section~\ref{section:desired_future_conceptualization} and present its outcomes in Section~\ref{section:desired_future_outcome}. We build on these results and qualitatively assess the progression towards the principles' ideals of the collectively determined DF and repeat this assessment for the status quo and inclination of trends. This combined evaluation in Section~\ref{section:towards_responsible_qi} allows us to make a normalized comparison between now and the desired long-term future, opening up the discussion on the direction of the ELSPI in achieving a responsible future QI.

\section{Desired future conceptualization}
\label{section:desired_future_conceptualization}
Pioneered by Pierre Wack at Royal Dutch Shell, the art of `Scenario Planning' brings a structured approach to develop possible futures and strategies based on them~\cite{wack1985scenarios}. This approach recognized that for successful scenario planning different points of view are crucial to prevent group thinking. Stakeholders also have mental models of the future, which have to be actively challenged, leading to re-interpretation of reality. We extend the scenario planning exercise with an explicit expression of the stakeholders' favorable future~\cite{schultz1995futures}. The complete steps of the methodology are:

\begin{enumerate}[noitemsep]
    \item Setting the focal topic
    \item Exploration of topic's influential factors (horizon scanning)
    \item Ranking these factors by impact and uncertainty
    \item Generate extreme scenarios representing two key uncertainties 
    \item Build these scenarios focusing on the implications they have on the topic
    \item Explicit articulation of a DF extracted from (4-5)
\end{enumerate}
For this paper the focal topic has been set to the `ELSPI of the QI', and in the next sections we go more into detail of steps 2-5. 

\subsection{Horizon scanning}
In order to fulfill step 2 of the methodology a systematic identification, monitoring and examination of relevant elements surrounding the topic has been performed, known as horizon scanning~\cite{Greely2022}. This scanning starts with identifying signals of new developments within the scope of the topic that potentially could influence the future in a broad sense. Signals are perceived indicators of small or local phenomena that have the potential to grow in scale and reach~\cite{signal_scanning_2022}. Horizon scanning therefore covers as many domains as possible, from politics, law, policy, natural environment and economy to technology and its impact on society. In total close to 200 signals have been found across all domains that surround the topic of the QI, which have been grouped into 14 `seeds of the future' based on theme and collective fundamentals, finalizing step 3.

\subsection{Scenario building}
\begin{figure}[t]
    \centering
    \includegraphics[width =0.77\linewidth]{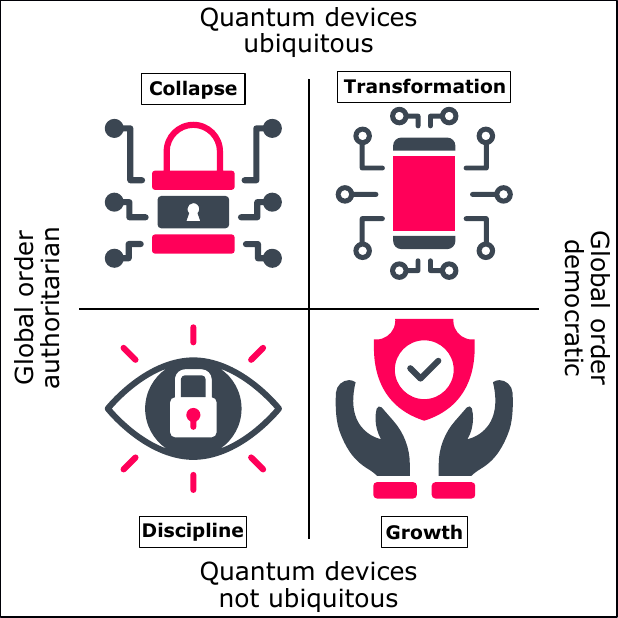}
    \caption{The chosen scenario logics matrix of extreme opposites of uncertain outcomes of the future. The scenarios are used in a world building exercise from which we can extract a DF. The four scenarios follow Dator's archetypes of generalized futures. }
    \label{fig:logics_matrix}
\end{figure}

To spark discussion and expand the perspectives of the stakeholders, two future uncertainties were carefully selected to form a `scenario logics matrix' (step 4), shown in Fig.~\ref{fig:logics_matrix}. The two axes of this matrix must be uncorrelated and opposite ends of both axes are extreme opposite outcomes of the uncertainties. 

The act of world building in these polarized scenarios has the power to extract underlying assumptions, break group thinking and form the `likes' and `dislikes' about each of the scenarios. The four outcomes align to Dator's archetypes of generalized futures~\cite{dator2019alternative} as shown in Fig.~\ref{fig:logics_matrix}. They describe either the continued growth of the current situation (Growth) or for the situation to completely inverse (Collapse). Additionally, we can describe a world where radical changes occur in all directions (Transformation) or scarcity prevails and rules and (environmental) regulations are prioritized to enable human survival (Discipline).

We continue with step 5, where stakeholders developed each of these scenarios from the point of view of The Netherlands in a European context. In all scenarios the implications of future applications of the QI were explored~\cite{Wehner2018}. Although this world building process aims to be as unbiased as possible, we need to recognize that all participants are currently living in a democratic society. This leads to an inherent sense that an authoritarian regime is not desirable and a democratic order is desired. Additionally, all participants are part of the quantum ecosystem, which makes it likely to inherently favor the success of the QI. Even though during this process it was stressed that participants should be as objective as possible in their world building, it is impossible to claim that this has occurred completely without bias. 

After this session, almost 40 participants in eight parallel groups of diverse composition explored in further detail the ELSPI of the QI in the Netherlands situated in the four scenarios. Each group was assigned to assess the implications of the QI either in the energy or healthcare sector, as example domains that play a critical part in society that combines technology with people's well-being. Then, groups exercised the ELSPI of the QI in their worlds, its (un)intentional benefits in their assigned scenario, also taking the opposing stance on the negative effects that QI applications can pose within these domains\footnote{Criminal activity being undetectable, dependence on governing entities to provide secure transmissions, quantum secure connections used to obfuscate means to control the population, etc.}. Lastly, they were encouraged to make recommendations for preemptive actions in the quantum network sector to be taken in 2024 to steer their scenario into a direction that was acceptable to them. This concludes step 5 of the process to create a DF.

\begin{figure*}[ht!]
    \centering
    \includegraphics[width = 0.86\textwidth]{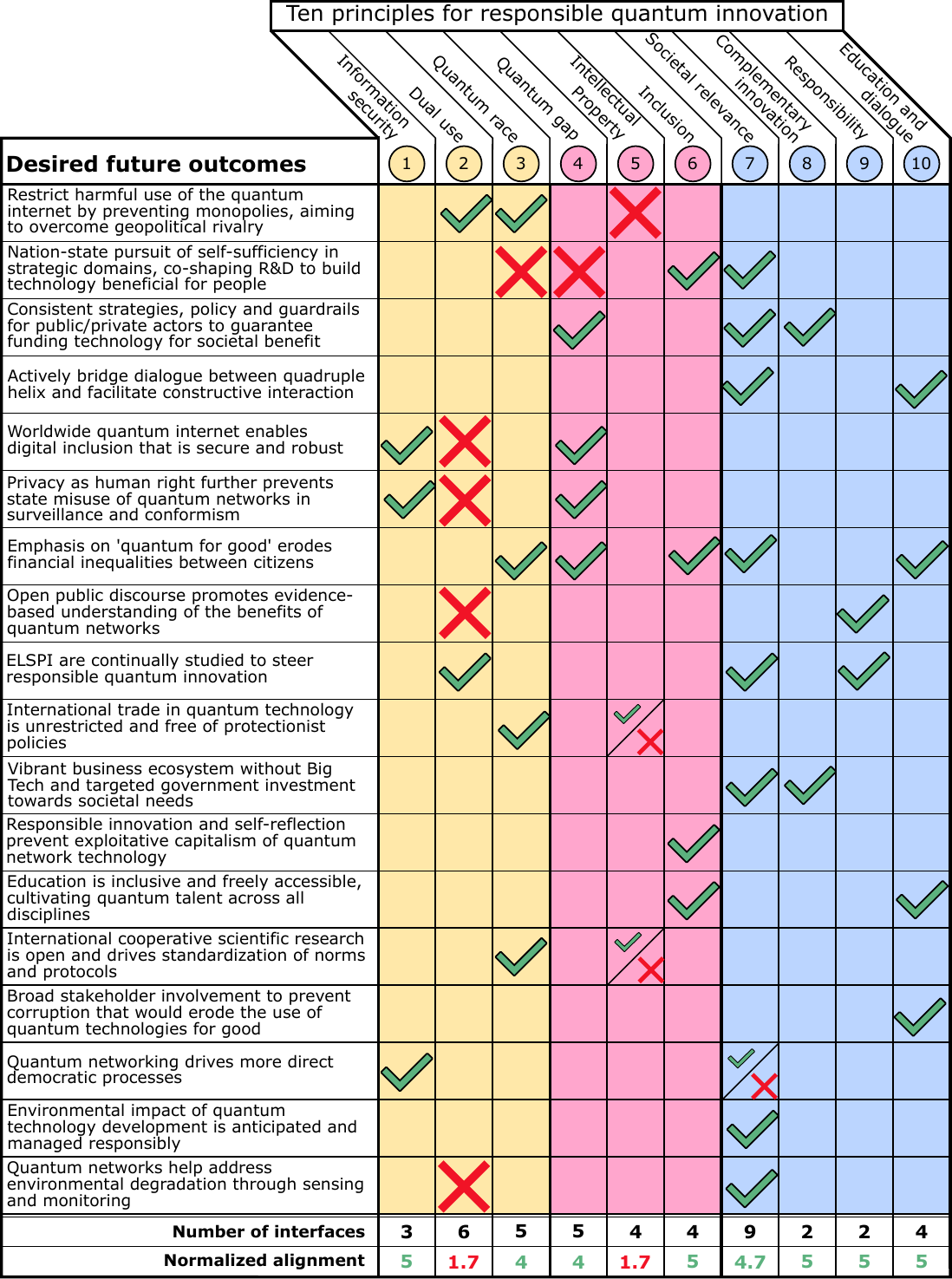}
    \caption{Mapping of the desired future conceptualization outcomes if they are `aligned' (\checkmark), `misaligned' ({\bfseries\large$\boldsymbol{\times}$}) or unrelated (empty) to each of the ten principles for responsible quantum innovation. The amount of interfaces is (\#\checkmark +\#{\bfseries\large$\boldsymbol{\times}$}). The alignment per principle is the sum (\#\checkmark - \#{\bfseries\large$\boldsymbol{\times}$}) normalized to the amount of interfaces of that principle and rescaled to be between $\left[0,5\right]$. This allows us to compare the alignment per principle, independent of the amount of interfaces of that principle in the DF. This normalized alignment is the input for the DF radial spokes of Fig.~\ref{fig:spider}. The split (\#\checkmark/ \#{\bfseries\large$\boldsymbol{\times}$}) signifies an alignment or misalignment dependent on their interpretation, which counts as `neutral' for the alignment and as a single interface. It is included to acknowledge the existence of an interface but one that is ambiguous or double-edged.}
    \label{fig:desired_future_responsible}
\end{figure*}

\begin{figure*}[t]
    \centering
    \includegraphics[width = 0.75\textwidth]{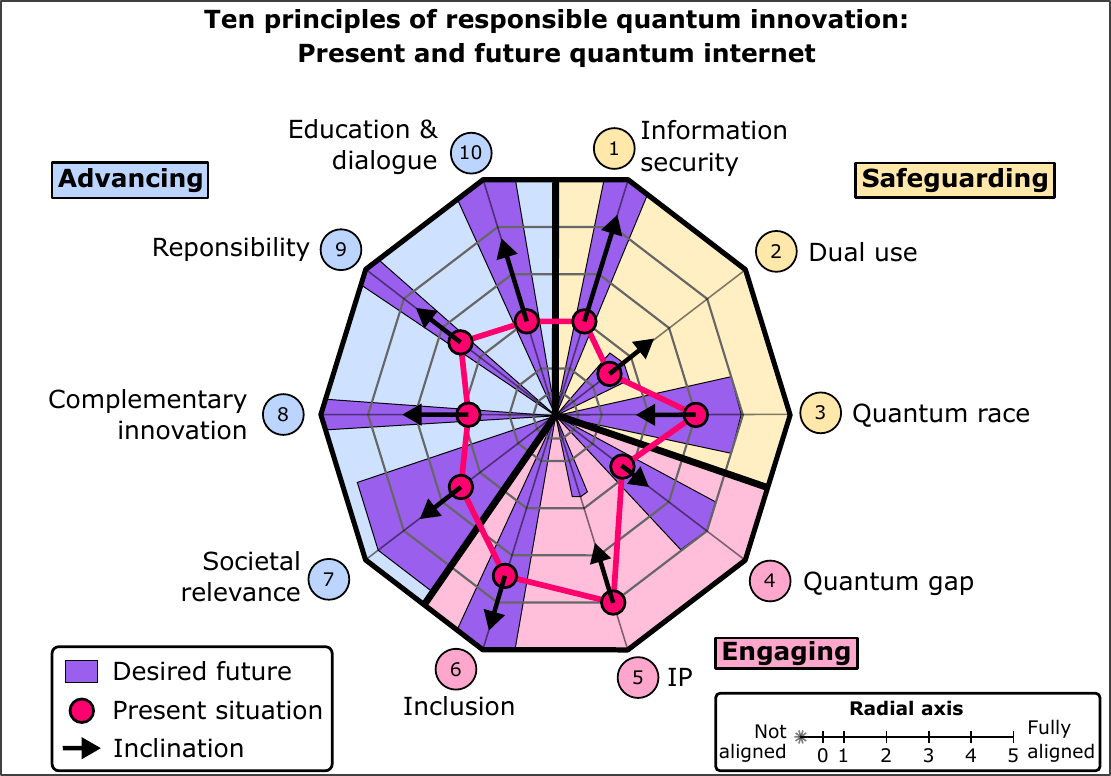}
    \caption{Mapping of the status of the development of a responsible QI for the present situation (pink dots, line), where we expect the trend to incline towards (black arrow) and for the DF outcomes (purple spokes) from Fig.~\ref{fig:desired_future_responsible}. The radial axis denotes alignment to the principle, with full alignment on the outer rim and `not aligned' on the inner most rim. The normalized interface occurrence of each principle in the DF is represented by the width of the spoke. The present situation alignment is determined through the authors' expert opinion, see Section~\ref{section:present_situation}.}
    \label{fig:spider}
\end{figure*}
\section{Desired Future}
\label{section:desired_future_outcome}
From the guided discussions and topics presented by the participants, we process a list of 18 outcomes of what collectively is framed to be the DF of the ELSPI of the QI, shown in Fig.~\ref{fig:desired_future_responsible}. These outcomes are mapped in `alignment' to each of the ten principles in Fig.~\ref{fig:desired_future_responsible}. We take each principle's definition and detailed description and map whether or not the DF outcome is aligned to the ideal that the principle describes. This assessment is the authors' expert opinion. Each DF outcome can have alignment with some and misalignment for other principles, simultaneously. As example we take the DF outcome `worldwide quantum internet enables digital inclusion that is secure and robust': With the push for information security inherent in this statement it is aligned to P1, and with expressing desire to give quantum internet access to everyone, it is aligned to P4. However, with unconstrained and global access to quantum internet facilities, malicious activity could secretly take place over secure communication channels, which is in misalignment to the prevention of dual uses of P2. We find the other principles not relevant to this DF outcome, and this finalizes the mapping for this outcome.

The normalized alignment of Fig.~\ref{fig:desired_future_responsible} is shown on the radial axis in Fig.~\ref{fig:spider}, where the width of the spokes is the normalized number of interfaces of the principle in the DF. We can interpret Fig.~\ref{fig:spider} as a measure of two parameters. First, the `balance' of the importance placed on each principle in the DF, where an ideal future gives all ten principles equal attention (equal and widest spoke width). Even though it cannot be expected that all topics that are covered in the workshop sessions would have equal representations, they are aimed to be equal enough that any topics that participants would over- or under-represent, will stand out. Second, the alignment of the principle in the DF to the ideal that the principle represents (radial axis). In the DF crafted by the stakeholder participants, we recognize that all principles are well aligned to the ideal except for `dual use' (P2) and `intellectual property' (P5).

\renewcommand{\thesubsection}{P\arabic{subsection}}
\section{Present situation and inclination}
\label{section:present_situation}
Alongside the DF ideation, we show an assessment of the `present situation' of the alignment with the ten principles in Fig.~\ref{fig:spider}, including an inclination of the alignment in the near-term future given the trends relevant at the time of writing of this paper. The placement on the alignment scale is determined by the authors' expert opinion and argumented by recent research papers and ELSPI developments. Since QI technology is still in development and fluid in directionality, we recognize that generally all ten principles are not fully developed yet, however the current assessment is a best-effort approach to provide a snapshot of the status quo. This present situation assessment as well as the following analyses are directly the authors' work and not developed in workshops.

\section{A responsible future quantum internet}
\label{section:towards_responsible_qi}
The overview of Fig.~\ref{fig:spider} presents the complete status of the present situation, future inclination and DF of the ten principles of responsible quantum innovation. We observe that the conceived DF and present situation have under-representation of interface occurrence or misalignment with several principles, which we discuss below.

\subsection{Analysis} 
A possible explanation of misalignment of the `dual use' principle (P3) both in the DF and present situation is the perceived fear that `negative' connotation to QI technologies is not advised to be publicly discussed, as to not harm the `positive' reputation the QI has in delivering enhanced security. We observe that this results in DF outcomes that are misaligned with P3. The same argument holds for the present situation: it is not a `positive' narrative to discuss misuse of quantum technologies other than to protect against adversarial use of those same technologies. With history showing to prepare for misuse of any new technology that has potential for destructive power, the QI domain requires this as well.

A different type of explanation can be found for the misalignment of the `IP' principle (P5) in the DF, where outcomes were formulated in favor of unrestricted and free of protectionist trade policies. This endangers the implementation of methods to protect a market if new risks are feared to destabilize a healthy economy. With one of the desired outcomes being to `restrict harmful use of the QI by preventing monopolies' (see Fig.~\ref{fig:desired_future_responsible}) we risk closing the possibility for large technology players to rise up from the average. Although clearly with its flaws~\cite{foer2018world, Hendrikse2022}, the existence of `Big Tech' does open up the possibility for enormous private R\&D spending (\euro140 billion in 2023 combined of the top 5 R\&D spenders~\cite{RD_EU_2023}) and the development of market-driven standardization that is much harder to achieve in an otherwise fragmented sector~\cite{petit_2020}. Preemptively decimating the possibility of large-cap QI companies risks missing out on developing a world-leading quantum industrial sector. As such, a balance of these topics needs to be found to remain aligned with P5.

In the present situation however the public discourse on P5 is more balanced. This is a consequence of the lower TRL of the QI domain, where most knowledge development occurs at research institutes, allowing for more transparency, collaboration and open research. With the TRL of the domain increasing over time, commercialization efforts can result in more secrecy and proprietary information~\cite{Kop2022}. Together with a current worldwide political push for sovereignty~\cite{Omaar2023, EU_foreign_investment_law, STEP2023}, relating to `quantum race' (P3), the future inclination of both principles tilts towards misalignment (inward facing arrows in Fig.~\ref{fig:spider}).

The `societal relevance' principle (P7) is overemphasized in occurrence in the DF conceptualization relative to the other principles (widest spoke in Fig.~\ref{fig:spider}). This is explained by the assumption that the QI domain should find (societal) useful applications as driver for technological development. This trend is in line with technological solutionism to solve global environmental challenges~\cite{saetra2023technology}, in which quantum technologies can potentially play a significant role~\cite{Dieterich2017,WEF_2019}. While this emphasis has good intentions, it has the potential side-effect of limiting technology research only for those purposes. The definition of P7 allows for advancing `base-layer' technologies~\cite{kop2023quantum}, however this is a nuance that stakeholders potentially oversee in the quest for solving humanity's global challenges. Additionally, a relative overemphasis of a principle risks removing attention from others, exemplified by the relative unbalanced representation of occurrences in the DF, should thus be avoided. 

\renewcommand{\thesubsection}{\Alph{subsection}}
\subsection{Points of future interest}
To improve on the balance of representation and alignment of several principles in the DF and present situation,  we make several recommendations for future discussions on the ELSPI of the QI:

\begin{enumerate}[itemsep=0mm, label=\textbf{(\arabic*)}]
    \item Create a critical review of currently known applications of the malicious use of QI technologies. This allows stakeholders to have an improved understanding of the risks involved and allows for ideation on prevention or enhanced safeguarding of dual use applications (P2).
    \item Closer involvement of (quantum) industry in QI ELSPI discussions. This opens the possibility to include a balancing voice on how industry views proposed policy and market regulation (P3, 5). 
    \item Actively emphasize the necessity of funding base-layer research that is not considered to be of direct societal relevance to prevent over-regulation of (commercial) research (P7).
    \item Educate stakeholders on the QI and expand to include more expert future stakeholders. The more stakeholders are aware of the field they will participate in, the better their steering towards the principles can be (P1, 8, 9, 10).
    
\end{enumerate}

\subsection{Limitations of assessment}
The DF conceptualization through workshops is bounded by its process as described in Section~\ref{section:desired_future_conceptualization} and the participants itself, which has its limitations. We suggest several process improvements. For more consistency and to allow for full understanding of the QI context, we suggest to have the same participants in every workshop round and to allot more time or sessions to fully cover all discussions. This allows for intermediate assessment and steering of principle representation. Furthermore, we suggest to expand the scope of the DF to include influences from other important technological trends (AI, biotech, sustainability) which are expected to have mutual interactions with quantum technologies, covering a more complete conceptualization within responsible innovation.

Lastly, although the DF is crafted through stakeholders, the principle alignment assessment of the DF and present situation is performed solely by the authors. This assessment can benefit from a broader perspective or involvement of principle-domain experts.

\section{Conclusion}
We have presented an evaluation of the development towards a responsible future QI, assessing both the status quo and a collectively conceptualized DF along the ideal of the ten principles of quantum innovation. Most principles in the DF are well aligned to the ideal, however we found there to be significant misalignment in the DF of principles related to `IP' and `dual use', revealing the precarious balance of well intended policy suggestions and effective outcomes. This misalignment is possibly driven by the unpopularity of discussing `negative' effects of quantum technologies, as well as an under-representation of commercial entity stakeholders in the DF ideation process. Additionally, there is an overemphasis on the principal of societal relevance in the DF, risking overseeing other principles in the future steering of the ELSPI. The present situation has on average moderate alignment with the principles, with most trends showing inclination in the direction of the ideal, except for the principles `IP' and `quantum race', driven by the trends of quantum technology commercialization and the intention of geopolitical sovereignty. 

For continued success of a responsible QI, we recommend further rounds of workshops with stakeholders with improved consistency and expanded technological scope. For the quantum ecosystem we recommend further investigation on the prevention of dual use QI applications, closer involvement of commercial entities in ELSPI ideation, continued recognition of base-layer technology research and stakeholder education on QI applications and developments.

\section*{Acknowledgment}
\noindent 
We thank all participants for their contributions to the `Scenarios for Quantum Networks' project, the expert group for their useful input and discussions and Prof. Deborah Nas for developing and leading the project.

\bibliographystyle{IEEEtran}
\bibliography{main_IEEE}

\begin{thebibliography}{10}
\providecommand{\url}[1]{#1}
\csname url@samestyle\endcsname
\providecommand{\newblock}{\relax}
\providecommand{\bibinfo}[2]{#2}
\providecommand{\BIBentrySTDinterwordspacing}{\spaceskip=0pt\relax}
\providecommand{\BIBentryALTinterwordstretchfactor}{4}
\providecommand{\BIBentryALTinterwordspacing}{\spaceskip=\fontdimen2\font plus
\BIBentryALTinterwordstretchfactor\fontdimen3\font minus \fontdimen4\font\relax}
\providecommand{\BIBforeignlanguage}[2]{{%
\expandafter\ifx\csname l@#1\endcsname\relax
\typeout{** WARNING: IEEEtran.bst: No hyphenation pattern has been}%
\typeout{** loaded for the language `#1'. Using the pattern for}%
\typeout{** the default language instead.}%
\else
\language=\csname l@#1\endcsname
\fi
#2}}
\providecommand{\BIBdecl}{\relax}
\BIBdecl

\bibitem{Kimble2008}
H.~J. Kimble, ``\BIBforeignlanguage{en}{The quantum internet},'' \emph{\BIBforeignlanguage{en}{Nature}}, vol. 453, no. 7198, pp. 1023--1030, Jun. 2008.

\bibitem{Degen2017}
C.~Degen, F.~Reinhard, and P.~Cappellaro, ``Quantum sensing,'' \emph{Reviews of Modern Physics}, vol.~89, no.~3, p. 035002, Jul. 2017.

\bibitem{Bova2021}
F.~Bova, A.~Goldfarb, and R.~G. Melko, ``\BIBforeignlanguage{en}{Commercial applications of quantum computing},'' \emph{\BIBforeignlanguage{en}{EPJ Quantum Technology}}, vol.~8, no.~1, p.~2, Dec. 2021.

\bibitem{Bayerstadler2021}
A.~Bayerstadler, G.~Becquin, J.~Binder, T.~Botter, H.~Ehm, T.~Ehmer, M.~Erdmann, N.~Gaus, P.~Harbach, M.~Hess, J.~Klepsch, M.~Leib, S.~Luber, A.~Luckow, M.~Mansky, W.~Mauerer, F.~Neukart, C.~Niedermeier, L.~Palackal, R.~Pfeiffer, C.~Polenz, J.~Sepulveda, T.~Sievers, B.~Standen, M.~Streif, T.~Strohm, C.~Utschig-Utschig, D.~Volz, H.~Weiss, and F.~Winter, ``\BIBforeignlanguage{en}{Industry quantum computing applications},'' \emph{\BIBforeignlanguage{en}{EPJ Quantum Technology}}, vol.~8, no.~1, p.~25, Dec. 2021.

\bibitem{Wehner2018}
S.~Wehner, D.~Elkouss, and R.~Hanson, ``Quantum internet: {A} vision for the road ahead,'' \emph{Science}, vol. 362, no. 6412, p. eaam9288, Oct. 2018.

\bibitem{kop2023quantum}
M.~Kop, ``{Quantum-ELSPI: a novel field of research},'' \emph{Digital Society}, vol.~2, no.~2, p.~20, 2023.

\bibitem{Kop2024}
M.~Kop, M.~Aboy, E.~D. Jong, U.~Gasser, T.~Minssen, I.~G. Cohen, M.~Brongersma, T.~Quintel, L.~Floridi, and R.~Laflamme, ``\BIBforeignlanguage{en}{Ten principles for responsible quantum innovation},'' \emph{\BIBforeignlanguage{en}{Quantum Science and Technology}}, vol.~9, no.~3, p. 035013, Apr. 2024.

\bibitem{Fuller2009}
T.~Fuller and K.~Loogma, ``Constructing futures: {A} social constructionist perspective on foresight methodology,'' \emph{Futures}, vol.~41, no.~2, pp. 71--79, Mar. 2009.

\bibitem{wack1985scenarios}
P.~Wack, ``Scenarios: uncharted waters ahead,'' \emph{Harvard business review}, vol.~63, no.~5, pp. 72--89, 1985.

\bibitem{schultz1995futures}
W.~L. Schultz, \emph{Futures fluency: explorations in leadership, vision, and creativity}.\hskip 1em plus 0.5em minus 0.4em\relax University of Hawai'i at Manoa, 1995.

\bibitem{Greely2022}
H.~T. Greely, ``\BIBforeignlanguage{en}{Governing emerging technologies—looking forward with horizon scanning and looking back with technology audits},'' \emph{\BIBforeignlanguage{en}{Global Public Policy and Governance}}, vol.~2, no.~3, pp. 266--282, Sep. 2022.

\bibitem{signal_scanning_2022}
\BIBentryALTinterwordspacing
G.~Profitiliotis, ``{HLCP} {Futures} {Masterclasses} {Pilot} 2022-2023: {Horizon} {Scanning},'' Dec. 2022. [Online]. Available: \url{https://www.youtube.com/watch?v=rN6I4dGHkb8}
\BIBentrySTDinterwordspacing

\bibitem{dator2019alternative}
J.~Dator, ``Alternative futures at the manoa school,'' \emph{Jim Dator: A Noticer in Time: Selected work, 1967-2018}, pp. 37--54, 2019.

\bibitem{foer2018world}
F.~Foer, \emph{World without mind: The existential threat of big tech}.\hskip 1em plus 0.5em minus 0.4em\relax Penguin, 2018.

\bibitem{Hendrikse2022}
R.~Hendrikse, I.~Adriaans, T.~J. Klinge, and R.~Fernandez, ``The {Big} {Techification} of {Everything},'' \emph{Science as Culture}, vol.~31, no.~1, pp. 59--71, Jan. 2022.

\bibitem{RD_EU_2023}
E.~Nindl, H.~Confraria, F.~Rentocchini, L.~Napolitano, A.~Georgakaki, E.~Ince, P.~Fako, A.~Tübke, J.~Gavigan, H.~Hernández~Guevara, P.~Pinero~Mira, J.~Rueda~Cantuche, S.~Banacloche~Sanchez, G.~De~Prato, and E.~Calza, ``The 2023 eu industrial r\&d investment scoreboard,'' 2023.

\bibitem{petit_2020}
N.~Petit, \emph{{Big Tech and the Digital Economy: The Moligopoly Scenario}}.\hskip 1em plus 0.5em minus 0.4em\relax Oxford University Press, 10 2020.

\bibitem{Kop2022}
\BIBentryALTinterwordspacing
M.~Kop, M.~Aboy, and T.~Minssen, ``Intellectual property in quantum computing and market power: a theoretical discussion and empirical analysis,'' \emph{Journal of Intellectual Property Law \& Practice}, vol.~17, no.~8, pp. 613--628, Aug. 2022. [Online]. Available: \url{https://doi.org/10.1093/jiplp/jpac060}
\BIBentrySTDinterwordspacing

\bibitem{Omaar2023}
\BIBentryALTinterwordspacing
H.~Omaar, ``\BIBforeignlanguage{en}{The {U}.{S}. {Approach} to {Quantum} {Policy}},'' \emph{\BIBforeignlanguage{en}{ITIF Center for Data Innovation}}, Oct. 2023. [Online]. Available: \url{https://www2.datainnovation.org/2023-us-quantum-policy.pdf}
\BIBentrySTDinterwordspacing

\bibitem{EU_foreign_investment_law}
``{Regulation (EU) 2019/452 of the European Parliament and of the Council of 19 March 2019 establishing a framework for the screening of foreign direct investments into the Union},'' \emph{OJ}, vol.~L, no. 79I, pp. 1--14, 2019.

\bibitem{STEP2023}
{European Commission}, ``{Proposal for a REGULATION OF THE EUROPEAN PARLIAMENT AND OF THE COUNCIL establishing the Strategic Technologies for Europe Platform (‘STEP’) and amending Directive 2003/87/EC, Regulations (EU) 2021/1058, (EU) 2021/1056, (EU) 2021/1057, (EU) No 1303/2013, (EU) No 223/2014, (EU) 2021/1060, (EU) 2021/523, (EU) 2021/695, (EU) 2021/697 and (EU) 2021/241},'' \emph{COM/2023/335 final}, 2023.

\bibitem{saetra2023technology}
H.~S. S{\ae}tra, \emph{Technology and sustainable development: The promise and pitfalls of techno-solutionism}.\hskip 1em plus 0.5em minus 0.4em\relax Taylor \& Francis, 2023.

\bibitem{Dieterich2017}
J.~M. Dieterich and E.~A. Carter, ``\BIBforeignlanguage{en}{Opinion: {Quantum} solutions for a sustainable energy future},'' \emph{\BIBforeignlanguage{en}{Nature Reviews Chemistry}}, vol.~1, no.~4, pp. 1--7, Apr. 2017.

\bibitem{WEF_2019}
J.~O'Brien, ``\BIBforeignlanguage{en}{How quantum computing could be one of the most innovative climate change solutions?}'' Dec. 2019.

\end{thebibliography}

\end{document}